\documentclass[12pt]{spieman}  
\usepackage{amsmath,amsfonts,amssymb}
\usepackage{graphicx}
\usepackage{setspace}
\usepackage{tocloft}
\usepackage{lineno}

\usepackage{caption} 
\newcommand{\cmark}{\color{black}}
\newcommand{\umark}{\color{black}}
\newcommand{\arcs}{$^{\prime\prime}$}

\linenumbers

\title{Isotopic composition of cometary water and the origin of Earth's oceans}

\author[a,*]{Dariusz C. Lis}
\author[b,c]{Martin Cordiner}
\author[d]{Nicolas Biver}
\author[d]{Dominique Bockel\'ee-Morvan}
\author[a]{Paul F. Goldsmith}
\author[e]{Arielle Moullet}
\author[a]{Paul von Allmen}

\affil[a]{Jet Propulsion Laboratory, California Institute of Technology, 4800 Oak Grove Drive, MS 169-506, Pasadena, CA 91109, USA}
\affil[b]{Department of Physics, Catholic University of America, Washington, DC 20064, USA}
\affil[c]{Solar System Exploration Division, Astrochemistry Laboratory Code 691, NASA-GSFC, Greenbelt, MD 20771, USA}
\affil[d]{LIRA, Observatoire de Paris, Universit\'{e} PSL, CNRS, Sorbonne Universit\'{e}, Universit\'{e} Paris Cit\'{e}, CY Cergy Paris Universit\'{e}, 5 place Jules Janssen, 92190 Meudon, France}
\affil[e]{National Radio Astronomy Observatory, 520 Edgemont Road, Charlottesville, VA 22903, USA}

\nolinenumbers

\cftpagenumbersoff{figure}
\cftpagenumbersoff{table} 
\begin{document} 
\maketitle

\begin{abstract} Studies of the water content and isotopic composition of water-rich asteroids and comets are of key interest for understanding the late accretion stage of the Solar System cometary and chondritic materials. \cmark The PRobe far-infrared Mission for Astrophysics (PRIMA) \umark can make an important contribution to solving this long-standing problem by carrying out direct measurements of the D/H ratio in a significant sample of Oort cloud and Kuiper belt comets, sampling the isotopic composition of the present-day outer Solar System. This would allow comparisons between different comet reservoirs, and with inner Solar System measurements in meteorites, as well as searching for correlations with physical parameters, such as hyperactivity, providing quantitative constraints on the dynamical and chemical models of the early Solar System.  \end{abstract}


{\noindent \footnotesize\textbf{*}D. C. Lis,  \linkable{Dariusz.C.Lis@jpl.nasa.gov} }

\begin{spacing}{2}   

\noindent $\copyright 2025$~All rights reserved

\section{Introduction}
\label{sect:intro}  

Understanding how the Earth obtained its water and whether water-rich Earthlike planets are common in the Universe is one of the central themes in NASA’s vision. The Decadal Survey \emph{Pathways to Discovery in Astronomy and Astrophysics for the 2020s} speaks of the core concept of gas and dust particles in a circumstellar disk assembling into planets. Our Solar System is currently the only laboratory where physical and chemical processes occurring in such disks and their impact on the habitability of planets forming in habitable zones can be investigated in detail. Understanding the trail of water from its interstellar reservoir to forming planetary systems (Figure 3 of Ref. \cite{Broadley22}) can be best studied by following the isotopic composition of water. An important fingerprint is the D/H isotopic ratio, which is strongly dependent on the water formation temperature and exhibits large variations among interstellar and Solar System objects. \cmark The required accuracy of individual measurements ($\sim 5 - 10$\%) is thus much less stringent compared to, e.g., measurements of the oxygen isotopic ratios in water, and achievable from remote sensing, as demonstrated by past observations\cite{Hartogh11, Bockelee12}. \umark Comparative studies of the water content and isotopic composition of protoplanetary disks \cite{Tobin23} and \cmark water-rich asteroids\cite{Greenwood23} and comets\cite{Hoppe23} are of key interest for understanding the late accretion stage of the Solar System cometary and chondritic materials. \umark

The D/H ratio in water can be measured from direct observations of water isotopologues in the far-infrared (FIR) and submillimeter through rotational spectroscopy, and in the near-infrared (NIR) through rovibrational fluorescence spectroscopy. It can also be derived indirectly from observations of water photodissociation products, H and OH, in the ultraviolet (UV). Both types of observations have been challenging in the past, but PRIMA\cite{Glenn25} can make an important contribution to solving the long-standing puzzle of the origin of Earth’s water by carrying out measurements of D/H in a significant sample of Solar System comets, contributing to the ultimate goal of obtaining a sample comparable in size to the current sample of measurements in meteorites. Comparing such measurements with observations of star-forming regions and now with ALMA observations of protoplanetary disks \cite{Tobin23} will allow us to test and quantitatively compare inheritance (water in forming planetary systems is directly inherited from its interstellar reservoir) and reset (ices are thermally and chemically reprocessed during the disk phase) hypotheses. Such observations will thus provide quantitative constraints for the dynamical/chemical models of protoplanetary disks and the early Solar System, vastly improving our understanding of the delivery of water and organic material to planets in habitable zones.

\section{Deuterium-to-hydrogen ratio in cometary water}
\label{doverh}

The trail of water from its interstellar reservoir to forming planetary systems can be best studied by following the isotopic composition of water. The D/H ratio strongly depends on the water formation temperature and exhibits large variations compared to, e.g., the oxygen isotopic anomalies \cite{Sakamoto07, McKeegan11}. The latest isotopic measurements in meteorites and models suggest that volatiles on the Earth and other terrestrial planets have been heterogeneously sourced from different solar system reservoirs \cite{Broadley22, Woo23}. Understanding the water content and isotopic composition of water-rich asteroids and comets is of key interest for understanding the late accretion stage \cite{vanDishoeck14}.

The outer asteroid belt is a natural reservoir of icy bodies in the Solar System. Models show that buried ice can persist within the top few meters of the surface of an asteroid over billions of years \cite{Schorghofer08} and Dawn observations \cite{Raponi18} showed that water ice can accumulate in permanently shaded areas of craters. Main-belt comets \cite{Hsieh06} constitute a separate class of intrinsically icy bodies in the outer asteroid belt and their water content was only recently confirmed using JWST \cite{Kelley23}.

Comets have formed and remained for most of their lifetime at large heliocentric distances. Therefore, they contain some of the least-processed, pristine ices remaining from the Solar Nebula disk\cite{Mumma11}, the composition of which can be studied through \cmark in situ measurements\cite{Krankowsky86} and remote sensing\cite{Crovisier99}. \umark Numerous complex molecules have been identified in comets from ground-based millimeter-wave remote sensing observations\cite{Biver19} and from \emph{Rosetta} observations of comet 67P\cite{Altwegg17b}. The derived composition of cometary ices is similar to low-mass star-forming regions, suggesting formation in the pre-solar cloud or in the cold outskirts of the Solar Nebula \cite{Drozdovskaya19, Ceccarelli23}.

\begin{figure}
  \begin{center}
    \begin{tabular}{c}
      \includegraphics[width=0.8\textwidth]{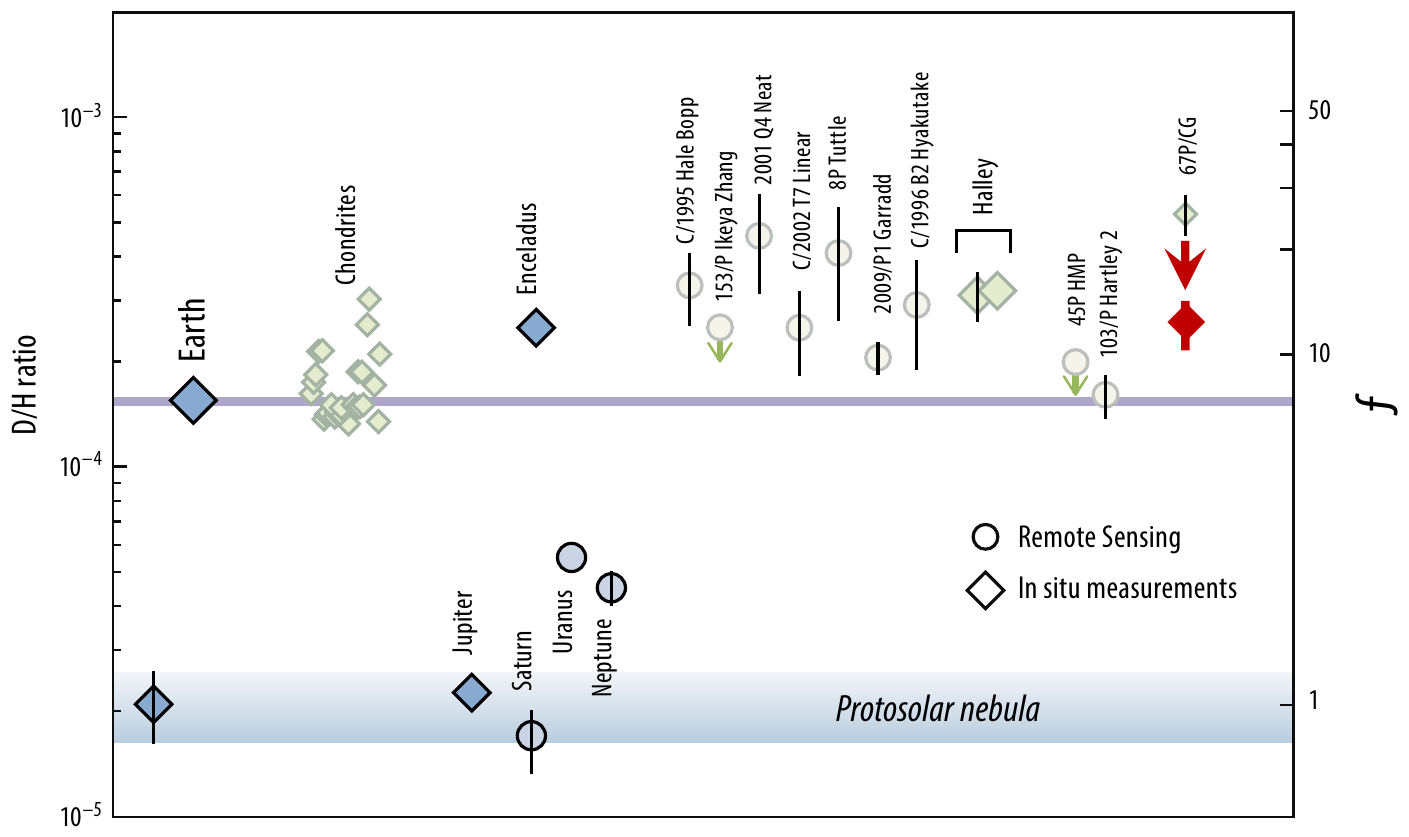}
    \end{tabular}
  \end{center}
  \caption
  { \label{fig:doverh} D/H ratio in Solar System objects.\cmark \cite{Meixner19, Altwegg15} \umark Comets exhibit large variations between 1 and 3 times Vienna Standard Mean Ocean Water (VSMOW). The red diamond and arrow mark the revised value of \cmark $2.59 \pm 0.36 \times 10^{-4}$ \umark in comet 67P/C-G\cite{Mandt24}.}
\end{figure}

Cometary D/H ratios span a significant range (Figure~\ref{fig:doverh}; from Ref.\cite{Meixner19}, based on Ref.\cite{Altwegg15}) explained by significant mixing of bodies from different regions in the Solar System caused by giant planet migration. It has also been suggested that a substantial fraction of Oort Cloud comets might have been captured from other stars in Sun’s birth cluster \cite{Levison10}. The variations in D/H ratio are currently interpreted as being the result of isotopic exchanges with D-poor gases (H$_2$, OH) in the hot, inner regions of the young solar nebula, combined with turbulent mixing \cite{Ceccarelli14, Nomura23}.  However, the high D$_2$O/HDO ratio measured in comet 67P suggests that cometary water is largely unprocessed and formed on grains in the presolar cloud \cite{Altwegg17a}.

While many comets show D/H ratios considerably in excess of the terrestrial value, a category of hyperactive comets has been identified in which the D/H ratio is consistent with the Earth’s value \cite{Lis19}. \cmark Hyperactive comets, typically defined as those having a nucleus active are fraction greater than 0.5, \umark contain icy grains expelled from the nucleus that sublimate completely in the coma and may be more representative of the bulk isotopic composition of the nucleus than water sublimating from the surface, which may be subject to stronger fractionation effects.

Observations to date, carried out over the past 35 years using ground-based and space facilities, have produced \cmark a dozen measurements \umark of the D/H ratio in comets, of which only 4 space-based measurements (comets 1P/Halley, 103P/Hartley 2, 67P/C-G, and 2009/P1 Garradd) have small enough uncertainties to be described as accurate \cmark (measurement uncertainty of 10\% or better, compared to $\sim 30$\% for ground-based measurements; see \umark Figure~\ref{fig:doverh}). We note that the highest D/H ratio in cometary water initially reported in comet 67P/C-G\cite{Altwegg15} has recently been questioned, as reanalysis of water isotope measurements over the full \emph{Rosetta} mission suggests that dust markedly increases the D/H ratio. The revised isotopic ratio in comet 67P/C-G measured at a distance from the nucleus where the gas is well mixed is in fact consistent with measurements in other Jupiter family comets\cite{Mandt24}. \cmark The new 67P/C-G analysis is applicable to PRIMA, which through remote sensing will primarily sample the well-mixed gas at large distances from the nucleus (the FIRESS Band 3 pixel size of 12.7\arcs\ corresponds to 9,200~km at 1~au). \umark

In spite of significant efforts to date, cometary D/H measurements are still very limited compared to over 100 measurements in meteorites, which probe the isotopic composition of the objects in the Asteroid belt\cite{Robert03}.  The long-term science objectives for studying the origin of the Earth’s water with PRIMA are to: (a) constrain the shape of the D/H probability density function in cometary water with accuracy comparable to the meteoritic measurements; (b) study differences in the D/H distribution between different dynamical classes of comets; (c) study correlations with physical parameters, such as hyperactivity. \cmark Observations of a large sample of comets with PRIMA, compared with similar measurements in disks and meteorites, will help improve our understanding of the complex water cycle in the Solar System, and by proxy in other exoplanetary systems. \umark

The statistical approach will allow addressing various testable hypotheses and narratives to describe how, when, and in what quantity comets contributed to the Earth’s water content. For example, if the Solar System small bodies that formed at different heliocentric distances in the protosolar disk were efficiently scattered due to the migration of giant planets, one may expect the same distribution of the D/H ratio in the Oort cloud and Kuiper belt objects. However, if a significant fraction of Oort cloud comets were captured from other stars in the Sun’s birth cluster, one may expect different distributions in the two dynamical reservoirs. PRIMA observations will thus provide quantitative constraints for the state-of-the-art dynamical/chemical models of the early Solar System.

\begin{table}
  \begin{center}
    
    \caption{Low-energy lines of water isotopologues in the PRIMA wavelength range.\\}\label{tab:lines}
    
\begin{tabular}{ccccccc}
\hline \hline 
  \rule[-3mm]{0mm}{8mm} Molecule & Transition & $\lambda (\mu$m) & $\nu$ (GHz) & log$_{10} A_{ij}$ & $E_l$ (K) & $E_u$ (K) \\
\hline 

HDO&	2(2,0)--1(0,1)&	106.65&	2810.88&	-3.613&	22.3&	157.2\\
HDO$^\dag$&	2(2,0)--1(1,1)&	125.85&	2382.17&	-0.887&	42.9&	157.2\\
HDO$^\dag$&	2(2,1)--1(1,0)&	130.84&	2291.31&	-0.897&	46.8&	156.7\\
HDO&	2(2,1)--2(0,2)&	159.35&	1881.29&	-3.914&	66.4&	156.7\\
HDO$^\dag$&	3(1,3)--2(0,2)&	184.44&	1625.41&	-1.348&	66.4&	144.4\\
HDO&	2(2,1)--2(1,2)&	196.85&	1522.93&	-1.685&	83.6&	156.7\\
HDO&	3(1,2)--2(1,1)&	198.90&	1507.26&	-2.182&	95.2&	167.6\\
HDO&	3(0,3)--2(0,2)&	221.45&	1353.78&	-2.276&	66.4&	131.4\\
HDO&	2(2,0)--2(1,1)&	232.10&	1291.64&	-1.797&	95.2&	157.2\\
HDO$^\dag$&	2(1,2)--1(0,1)&	234.64&	1277.68&	-1.658&	22.3&	83.6\\
HDO$^\ddag$&	1(1,0)--1(0,1)&	588.65&	509.29 &	-2.635&	22.3&	46.8\\
\hline						
H$_2^{16}$O&	2(2,0)--1(1,1)&	100.98&	2968.75&	-0.587&	53.4&	195.9\\
H$_2^{16}$O&	2(2,1)--1(1,0)&	108.07&	2773.98&	-0.594&	61.0&	194.1\\
H$_2^{16}$O&	3(0,3)--2(1,2)&	174.63&	1716.77&	-1.301&	114.4&	196.8\\
H$_2^{16}$O$^\ddag$&	2(1,2)--1(0,1)&	179.53&	1669.90&	-1.256&	34.2&	114.4\\
H$_2^{16}$O$^\ddag$&	2(2,1)--2(1,2)&	180.49&	1661.01&	-1.518&	114.4&	194.1\\
H$_2^{16}$O&	2(2,0)--2(1,1)&	243.97&	1228.79&	-1.731&	136.9&	195.9\\
H$_2^{16}$O$^\ddag$&	1(1,0)--1(0,1)&	538.29&	556.94 &	-2.465&	34.2&	61.0\\
\hline					
H$_2^{17}$O&	2(2,0)--1(1,1)&	101.52&	2952.96&	-0.592&	53.1&	194.9\\
H$_2^{17}$O&	2(2,1)--1(1,0)&	108.74&	2756.84&	-0.599&	60.7&	193.0\\
H$_2^{17}$O&	3(0,3)--2(1,2)&	174.49&	1718.12&	-1.294&	114.0&	196.4\\
H$_2^{17}$O&	2(1,2)--1(0,1)&	180.33&	1662.46&	-1.258&	34.2&	114.0\\
H$_2^{17}$O&	2(2,1)--2(1,2)&	182.09&	1646.40&	-1.526&	114.0&	193.0\\
H$_2^{17}$O$^\ddag$&	1(1,0)--1(0,1)&	543.08&	552.02 &	-2.473&	34.2&	60.7\\
\hline					
H$_2^{18}$O&	2(2,0)--1(1,1)&	102.00&	2939.00&	-0.600&	52.9&	193.9\\
H$_2^{18}$O&	2(2,1)--1(1,0)&	109.35&	2741.67&	-0.606&	60.5&	192.0\\
H$_2^{18}$O&	3(0,3)--2(1,2)&	174.37&	1719.25&	-1.291&	113.6&	196.2\\
H$_2^{18}$O&	2(1,2)--1(0,1)&	181.05&	1655.87&	-1.263&	34.2&	113.6\\
H$_2^{18}$O&	2(2,1)--2(1,2)&	183.53&	1633.48&	-1.536&	113.6&	192.0\\
H$_2^{18}$O$^\ddag$&	1(1,0)--1(0,1)&	547.39&	547.68 &	-2.483&	34.2&	60.5\\
\hline
\end{tabular}
\end{center}
Note: Entries in the table are: molecule, transition, wavelength ($\mu$m), frequency (GHz), log of the Einstein A coefficient, lower and upper level energies (K).
$^\dag$Strongest HDO lines that will be targeted by PRIMA. $^\ddag$Lines previously observed at high spectral resolution by \emph{Herschel}/HIFI.
\end{table}

\begin{figure}
\begin{center}
\begin{tabular}{c}
\includegraphics[width=0.8\textwidth]{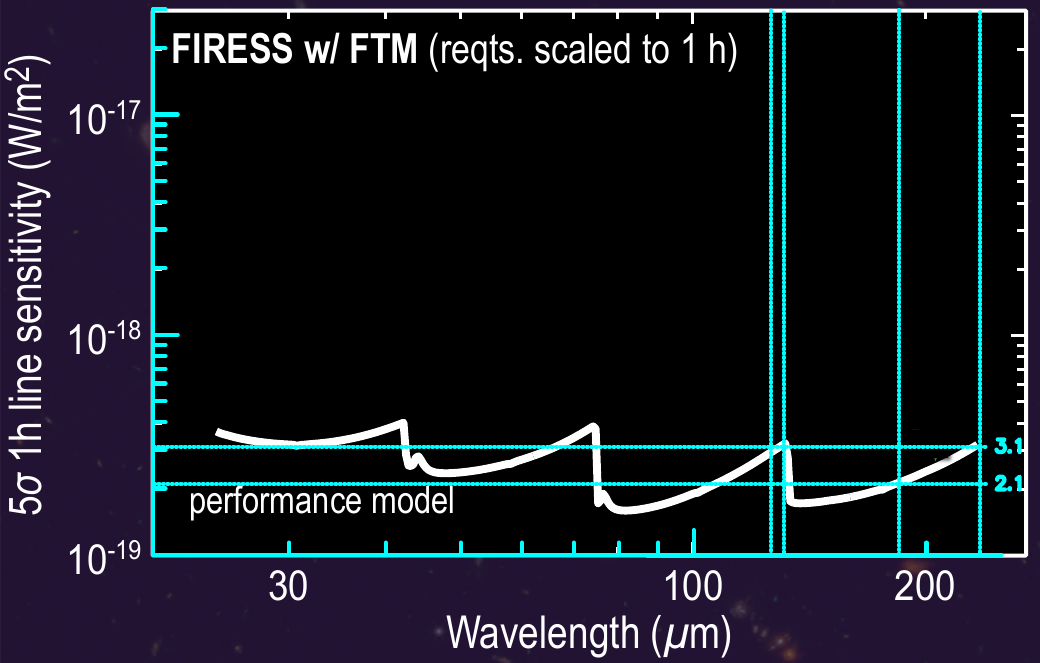}
\end{tabular}
\end{center}
\caption 
{ \label{fig:sensitivity}
Current best estimate (CBE) sensitivity of the FIRESS/FTM instrument as a function of wavelength (solid white line). Vertical dotted lines mark the wavelengths of the four brightest HDO lines. \cmark From the PRIMA Factsheet (https://prima.ipac.caltech.edu/page/fact-sheet). \umark} 
\end{figure}

\section{Deuterium-to-hydrogen ratio measurements with PRIMA}
\label{instreq}

PRIMA\cite{Glenn25} is a cryogenically cooled 1.8~m diameter space telescope. Given the weakness and narrow width of isotopic water lines in cometary atmospheres ($\sim 1$~km\,s$^{-1}$), a high spectral resolution is required to achieve a detection level required for isotopic ratio measurements. This implies the use of the FIRESS instrument in the FTM mode\cite{Bradford25}, which covers wavelength range from 24 to 235~$\mu$m with spectral resolution $R=4400 \times (112~\mu{\rm m}/\lambda)$, insufficient to spectrally resolve the narrow cometary lines, \cmark but sufficient to separate them from other molecular lines in cometary spectra to provide accurate integrated line fluxes. \umark  We use the current best estimate (CBE) FIRESS sensitivity \cmark from the Instrument Sheet \umark to determine how many D/H measurements can be carried out during the 5-year PRIMA primary mission. \cmark The FIRESS performance model is described in detail in in the accompanying instrument paper\cite{Bradford25}. The final sensitivity will likely evolve as the instrument design matures. We note, however, that for the SPIRE FTS instrument the in-orbit sensitivity was about a factor of 2 better that the pre-launch estimate. \umark We then consider two confounding factors that may affect observations of narrow spectral lines with a medium-resolution direct detection instrument: blending with nearby lines of other abundant coma species and the low expected line-to-continuum ratio.  For comparison, we also consider observations with a \cmark future passively-cooled ASTHROS\cite{Pineda22, Siles20}-like 2.5~m diameter FIR telescope \umark equipped with a state-of-the-art heterodyne instrument operating in the 500~GHz frequency band. Such observations are not impacted by these confounding factors and allow much longer integration times.

Since low-energy H$_2^{16}$O lines in cometary atmospheres are optically thick, accurate 3D radiative transfer modeling is required to accurately derive the  H$_2^{16}$O production rates. Consequently, the D/H ratio in water is determined in practice from observations of optically thin lines of the HDO and H$_2^{18}$O isotopologues \cite{Hartogh11, Lis19}.  This approach is justified as the $^{16}$O/$^{18}$O ratio exhibits much smaller variations in comets compared to the D/H ratio\cite{Biver07, Hartogh11, Paquette18}. Table~\ref{tab:lines} lists lines of water isotopologues with upper level energies below 200~K in the PRIMA wavelength range, which are potential targets. The 500~GHz lines previously used by \emph{Herschel}/HIFI for isotopic ratio measurements are also included. The strongest HDO lines at 125.85, 130.84, 184.44, and 234.64 $\mu$m are marked with a dagger in Table~\ref{tab:lines}. \cmark While the 234.64 $\mu$m HDO line is close to the upper end of the FIRESS wavelength range, the remaining lines are easily accessible. In particular, the 184.44 $\mu$m line is near the center of Band 4 of the spectrometer. \umark

Figure~\ref{fig:sensitivity} shows the FIRESS/FTM wavelength coverage. The 184.44 and 234.64 $\mu$m HDO lines can be observed simultaneously in the fourth module of the spectrometer, together with the 174.37, 181.05, and 183.53~$\mu$m lines of H$_2^{18}$O \cmark and the 174.49, 180.33 and 182.09~$\mu$m lines of H$_2^{17}$O. \umark The 125.85 and 130.84~$\mu$m HDO lines can be observed simultaneously with the 102.00 and $109.35$~$\mu$m lines of H$_2^{18}$O in the third module of the spectrometer. Two separate observations would be required to cover all the lines of interest.

\subsection{Line sensitivity}\label{sec:sensitivity}

The solid white line in Figure~\ref{fig:sensitivity} shows the current best estimate (CBE) FIRESS/FTM sensitivity as a function of wavelength.  Dashed vertical cyan lines mark the wavelengths of the four brightest HDO lines. The expected line sensitivity ($5 \sigma$, 1 h) is $2.1 \times 10^{-19}$ W\,m$^{-2}$ at 184.44  $\mu$m and about $3.1 \times 10^{-19}$ W\,m$^{-2}$ at the wavelengths of the other three HDO lines.

To estimate the sensitivity of a hypothetical far-infrared heterodyne space instrument for HDO detection in comets, we use archival \emph{Herschel}/HIFI observations of comet 103P/Hartley 2 \cite{Hartogh11}. These observations were carried out in the frequency switching (FSW) mode and consist of 11 scans of the H$_2^{18}$O 547.7~GHz line of 491 s duration each (on source time of 360 sec) and 10 scans of the HDO 509.3~GHz line of 2040~s duration each (on source time 1920 s). The observations were carried out using the High-Resolution Spectrometer (HRS) spectra with a frequency channel width of 0.12~MHz (velocity channel width of 0.066 and 0.070 km\,s$^{-1}$ for H$_2^{18}$O and HDO, respectively). H and V instrumental polarizations were averaged together doubling the integration time. The resulting \emph{Herschel} line sensitivity is 14.9 mK~kms$^{-1}$ ($5 \sigma$, 1 h on source time; main beam brightness temperature scale). We use the K\,kms$^{-1}$ units for the 509~GHz line flux rather than W\,m$^{-2}$, as these are standard for FIR/submm heterodyne instruments. We assumed a HIFI main beam efficiency of 62\%, as given in Ref.\cite{Shipman17}. for observations of HDO at 509~GHz. With frequency-switching, the on-source time will be close to the total observing time. Alternatively, a two-pixel instrument would allow position switching observations, which gives much flatter baselines, without doubling the total observing time to observe the reference position.

We note that the system temperature of the current state-of-the-art heterodyne instruments in the 500 GHz frequency band offers a factor of $\sim 1.3$ sensitivity improvement over HIFI. In addition, HIFI spectra had a 1.52 loss factor in the 2 bit, 3-level HRS correlator, and an additional factor of 1.2 sensitivity improvement can expected due to a higher main beam efficiency (75\% compared to HIFI's 62\%, Ref.\cite{Shipman17}). An overall sensitivity improvement by more than a factor of $\sim 2$ over the HIFI value reported in Table~\ref{tab:hdo} is thus expected in a future heterodyne mission, decreasing the observing time by a factor of 4. However, we use the more conservative HIFI value in the calculations below.

\begin{table}
\begin{center}  
\caption{Non-LTE HDO model line intensities for the reference coma model.}
\label{tab:hdo}
\begin{tabular}{cccccc}
\hline \hline 
  \rule[-3mm]{0mm}{8mm} Telescope & Transition & $\lambda$   & Line Flux & Sensitivity & Time\\
  \hline 
  PRIMA & 2(2,0)--1(1,1)&	125.85& $8.46 \times 10^{-20}$ & $3.1 \times 10^{-19}$ & 13.4 \\
  & 2(2,1)--1(1,0)&	130.84&	$8.32 \times 10^{-20}$ &  $3.1 \times 10^{-19}$ & 13.9 \\
  & 3(1,3)--2(0,2)&	184.44&	$8.23 \times 10^{-20}$ &  $2.1 \times 10^{-19}$ & 6.5 \\
  & 2(1,2)--1(0,1)&	234.64&	$1.30 \times 10^{-19}$ &  $3.1 \times 10^{-19}$ & 5.7 \\
\hline
  ASTHROS & 1(1,0)--1(0,1)&	588.65&	4.6 mK\,kms$^{-1}$ & 14.9 mK\,kms$^{-1}$     & 10.5  \\
  \hline
\end{tabular}
\end{center}
Note: Entries in the table are Telescope, HDO transition, wavelength in $\mu$m, line flux and assumed line sensitivity (5~$\sigma$, 1~h) in W\,m$^{-2}$ (mK\,kms$^{-1}$ for the 588.65~$\mu$m FIR line), and observing time required to reach a S/N ratio of 5 in hours.
The reference model assumed a water production rate of $2\times10^{28}$ s$^{-1}$, geocentric and heliocentric distance of 1 au, and a D/H ratio 2 times VSMOW. We assumed a telescope size of 1.8~m for PRIMA and 2.5~m for ASTHROS.
\end{table}


Table~\ref{tab:hdo} gives expected HDO line intensities computed using a non-LTE radiative transfer code for a reference model with a water production production rate $Q=2 \times 10^{28}$~s$^{-1}$, geocentric and heliocentric distances $r_h = \Delta = 1$~au (Figure of Merit, defined as the water production rate divided by the distance in au, FOM~$=2 \times 10^{28}$), and a D/H ratio of 2 times VSMOW (average value for past measurements in comets). The last column gives the observing time required to reach a S/N ratio of 5, given the instrument sensitivities listed above.

The radiative transfer code \cite{Cordiner22} is similar to the model used by \cite{Bockelee12, Lis19} and follows the molecular excitation as function of radius in the outflowing coma gas, including collisions with H$_2$O and electrons, as well as spectral line cooling and radiative pumping by the Solar radiation field. The outflow velocity is set to 0.8 km/s, and the HDO photolysis rate is assumed to be the same as water \cite{Huebner15}.

\begin{table}
\begin{center}  
\caption{Abundances of molecular species included in PSG models.}
\label{tab:mixrat}
\begin{tabular}{ccc}
\hline \hline 
  \rule[-3mm]{0mm}{8mm} Molecule & $X_{min}$ & $X_{max}$ \\
\hline 

  CH$_3$OH & 0.6 & 6.2 \\
  H$_2$CO  & 0.13 & 1.4 \\
  HCN      & 0.08 & 0.25 \\
  HNC      & 0.002 & 0.035\\
  H$_2$S   & 0.13 & 1.5 \\
  SO$_2$   & 0.2 & 0.2 \\
  CO       & 0.2 & 23 \\
  HCOOH    & 0.028 & 0.18 \\
\hline
\end{tabular}
\end{center}
Note: Abundances relative to water in percent. From Ref.\cite{Bockelee17}.
\end{table}

\subsection{Line blending}\label{sec:blending}

To estimate the impact of line blending for cometary D/H measurement at the spectral resolution of the FIRESS/FTM instrument we generated LTE model spectra of a cometary atmosphere using the NASA \cmark Planetary Spectrum Generator (PSG) \umark \cite{Villanueva18}. The key parameters of the model are the water production rate, $Q$, the coma temperature $T$, and the molecular abundances with respect to water. For the reference model, we use the same parameters as above, a water production rate $Q=2 \times 10^{28}$~s$^{-1}$, geocentric and heliocentric distances $r_h = \Delta = 1$~au. The coma temperature near 1~au was determined to be in the range $40-60$~K from observations of multiple methanol lines in comets 46P/Wirtanen and 21P/Giacobini-Zinner \cite{Biver21a, Biver21b}. We use 40~K in our calculations, as the LTE model will likely overestimate intensities of highly excited lines of the possible contaminating species, which are sub-thermally excited under the physical conditions in the coma.

Molecules included in the calculations are listed in Table~\ref{tab:mixrat}. Molecular abundances can vary significantly among comets \cite{Bockelee17}. As a conservative assumption, we assume the highest observed values in the line blending calculations. Out of the species included in the calculations, methanol has the highest density of lines in the far-infrared and is the most important contaminant.

\begin{figure}
\begin{center}
\begin{tabular}{c}
\includegraphics[trim=2cm 1.8cm 2.5cm 5cm, clip=true, width=0.8\textwidth]{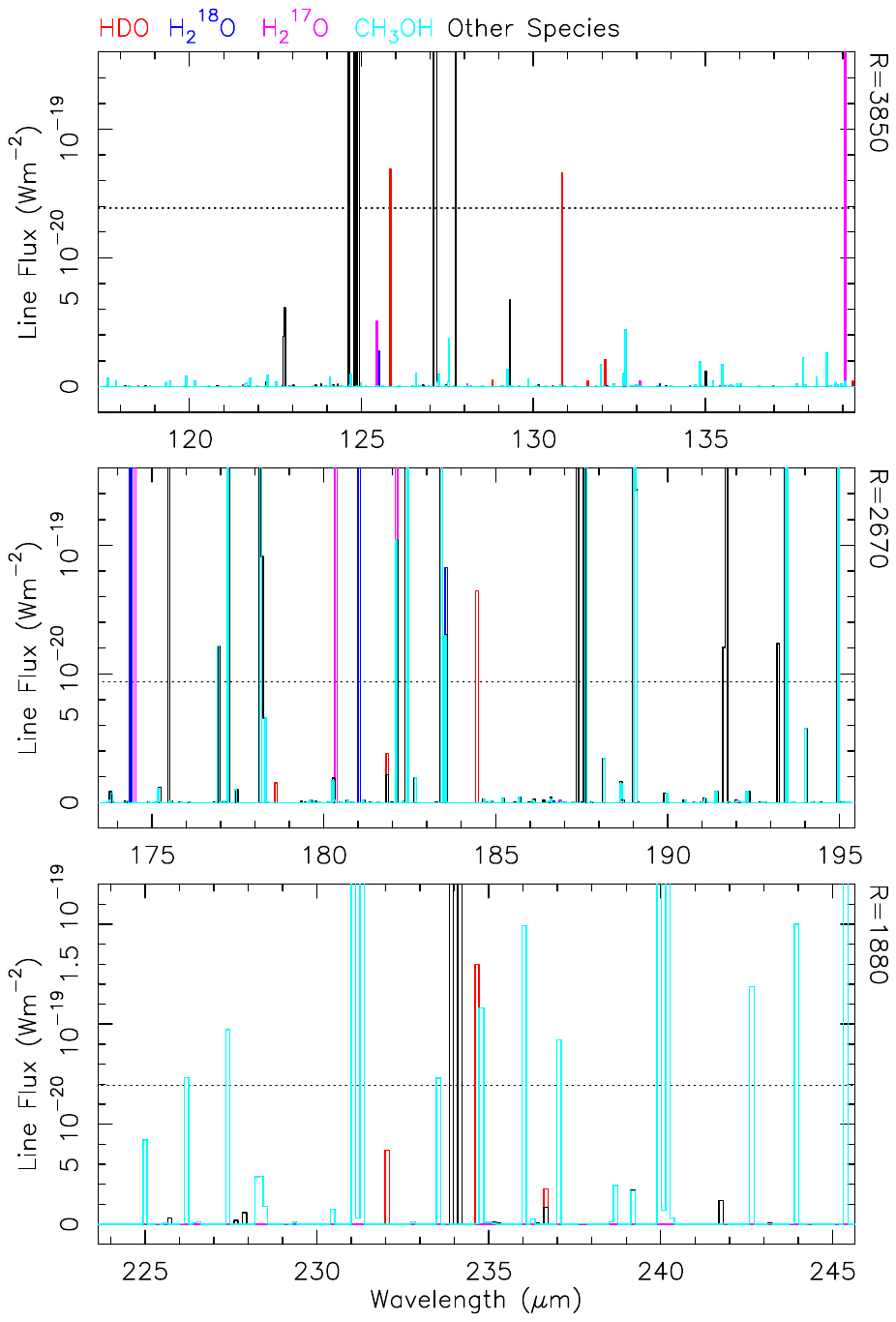}
\end{tabular}
\end{center}
\caption
{ \label{fig:contam} Molecular spectra for our reference coma model at the spectral resolution of FIRESS/FTM in the vicinity of the brightest HDO lines. HDO line fluxes (shown in red) correspond to the predictions of the full non-LTE excitation model, while those of the remaining lines are computed using the PSG tool, assuming a coma temperature of 40~K. Dotted horizontal lines mark expected $5 \sigma$, 20~h sensitivities of FIRESS/FTM.}
\end{figure} 

Figure~\ref{fig:contam} shows resulting PSG-generated spectra of the coma in the vicinity of the four brightest HDO lines in the PRIMA wavelength range. HDO lines are shown in red with line fluxes scaled to the predictions of the non-LTE model (Table~\ref{tab:hdo}). All other lines are as computed by the PSG LTE model. Lines of H$_2^{18}$O, H$_2^{17}$O, and methanol are shown in blue, magenta, and cyan, respectively, while lines of other abundant cometary specties are shown in black. Horizontal dotted lines mark the expected FIRESS/FTM sensitivity ($5 \sigma$, 20~h). The 125.85, 130.84, and 184.44~$\mu$m lines are not affected by blending with lines of other molecular species. \cmark The 184.44~$\mu$m HDO line, in combination with the nearby 181.05 H$_2^{18}$O and 180.33 H$_2^{17}$O lines, seems a particularly good target for isotopic ratio measurements, as our models show no evidence of blending with nearby lines of other molecular species.

\umark The brightest 234.64~$\mu$m HDO line is right next to a bright methanol line at 234.761~$\mu$m (one channel difference at $R=1880$). Blending with methanol thus has to be carefully considered in comets with high methanol abundance. However, as can be seen in Figure~\ref{fig:contam}, many methanol lines with comparable or higher line fluxes are present in the spectrum. \cmark In our PSG model, we count 21 such lines in Band 4 of the FIRESS spectrometer alone. These lines can be modeled carefully to predict the line flux of the methanol line that may be blended with the 234.64 HDO line. In addition to PSG models used here, which assume LTE excitation, a more sophisticated non-LTE models can be used in such analysis.\cite{Cordiner17, Biver21a, Biver21b, Cordiner22} The methanol line flux predicted by the model can then be subtracted to recover the unblended HDO line flux. \umark

In conclusion, line blending with other abundant coma species is not a confounding factor for D/H measurements in comets with PRIMA.

\subsection{Line to continuum ratio}\label{sec:lcr}

To estimate the expected line-to-continuum ratio for observations of HDO lines in comets using FIRESS/FTM we use archival 180 and 120~$\mu$m \emph{Herschel} PACS observations of comets 10P/Tempel~2, 103P/Hartley~2, and C/2009~P1 (Garradd) observed at 1--1.8 au from the Sun. We use spatially oversampled \emph{Rebinned} PACS data cubes (HPS3DRR), with a spaxel size of $8.6^{\prime\prime}$, from the \emph{Herschel} Science Archive (\emph{Herschel} OBSIDs 1342199882, 1342209391, 1342210191, 1342231307, 1342231308, 1342231309). To estimate the total continuum flux we sum up the 5 central spaxels with weights of 1.0 for the central spaxel and 0.5 for the offset spaxels. We verified that this approach gives fluxes consistent with published PACS observations of C/2006 W3 (Christensen)\cite{Bockelee10}. To compare the PACS observations with out reference model, we scale them to $r_h = \Delta = 1$~au and $Q= 2 \times 10^{28}$~s$^{-1}$ using the formula $S \sim Q \times \Delta^{-1} \times r_h^{-0.5}$, where the factor $r_h^{-0.5}$ accounts for the change in the dust temperature with the heliocentric distance. The resulting fluxes are listed in Table~\ref{tab:pacs}. The factor of $\sim 3$ scatter \cmark among different objects \umark is due to intrinsic variations among comets (e.g., differences in the dust-to-gas ratio, grain-size distribution etc.)

\begin{table}
\begin{center}  
\caption{\emph{Herschel} PACS observations of comets at 1--2 au.}
\label{tab:pacs}
\begin{tabular}{cccccccc}
\hline \hline 
  \rule[-3mm]{0mm}{8mm} Comet & $r_h$ & $\Delta$ & $Q$(H$_2$O) & O(180 $\mu$m) & O(120~$\mu$m) & S(180 $\mu$m) & S(120~$\mu$m) \\
\hline 
  10P/Tempel 2        & 1.43 & 1.80 & $2.2 \times 10^{28}$ & 0.18 & 0.71 & 0.35 & 1.38 \\
  103P/Hartley 2      & 1.07 & 0.18 & $1.2 \times 10^{28}$ & 0.42 & 1.44 & 0.13 & 0.45 \\
  103P/Hartley 2      & 1.09 & 0.21 & $1.2 \times 10^{28}$ & 0.60 & --   & 0.22 & -- \\
  C/2009 P1 & 1.78 & 1.93 & $2.0 \times 10^{29}$ & 0.69 & 3.17 & 0.18 & 0.82 \\
\hline
\end{tabular}
\end{center}
Note: Entries in the table are: comet, heliocentric and geocentric distances (au), water production rate (s$^{-1}$), observed 180 and 120~$\mu$m continuum fluxes (Jy), continuum fluxes scaled to the parameters of our reference coma model, as described in the text (Jy). The fluxes correspond to a weighted sum of the five central 8.6$^{\prime\prime}$ spaxels, as described in the text.
\end{table}

\begin{table}
\begin{center}  
\caption{Line to continuum ratio for observations of HDO using FIRESS/FTM.}
\label{tab:lcrat}
\begin{tabular}{ccccccc}
\hline \hline 
  \rule[-3mm]{0mm}{8mm} $\lambda$ & $R$ & $C_{min}$ & $C_{max}$ & $L$ & $L/C$ \\
\hline 
  120     & 3850 & $2.9 \times 10^{-18}$ & $8.9 \times 10^{-18}$ & $8.4 \times 10^{-20}$ & $0.9 - 2.9$ \\
  180     & 2670 & $8.2 \times 10^{-19}$ & $2.2 \times 10^{-18}$ & $8.2 \times 10^{-20}$ & $3.8 - 10.$ \\
\hline
\end{tabular}
\end{center}
Note: Entries in the table are: wavelength ($\mu$m), spectral resolution of FIRESS/FTM, minimum and maximum continuum fluxes in a spectrometer channel and HDO line flux (W~m$^{-2}$), line to continuum ratio (\%).
\end{table}

Table~\ref{tab:lcrat} gives the expected range of the continuum fluxes at 120 and 180~$\mu$m in a single channel of the spectrometer, as compared to the HDO fluxes computed above using the non-LTE excitation model. For the 125~$\mu$m HDO lines, our calculations suggest a line-to-continuum ratio of $\sim 1-3$~\%, near the expected lower limit for the line-to-continuum ratio achievable with a medium-resolution direct detection spectrometer, as demonstrated by past space instruments. The line-to-continuum ratio for the 184.44~$\mu$m HDO line is $\sim 4 - 10$~\%, and it is expected to be even higher for the 234.64~$\mu$m line, which is predicted to have higher line flux while the continuum flux decreases steeply with wavelength. We thus conclude that at the resolution of FIRESS/FTM the line-to-continuum ratio is not a confounding factor for measurements of the D/H ratio in comets using the 184.44 and 234.64~$\mu$m HDO lines. Since these two lines also offer the highest S/N ratio, we consider them preferred targets for D/H measurements in comets with PRIMA.

\section{Number of expected targets}\label{sec:numcomets}

We use the calculations from the \emph{Origins} Report\cite{Meixner19} (Figure 1--40) to estimate the number of comets accessible to PRIMA. For FOM~$= 2 \times 10^{28}$ (water production rate $Q = 2 \times 10^{28}$~s$^{-1}$ at $r_h = \Delta = 1$~au) we estimate that $17 \pm 2$ comets should be accessible during a nominal 5-year mission. For FOM~$=1 \times 10^{28}$, the number increases by a factor of $\sim 2$ to about 36.

\begin{figure}
\begin{center}
\begin{tabular}{c}
\includegraphics[trim=2cm 1.8cm 3.5cm 4.5cm, clip=true, width=0.8\textwidth]{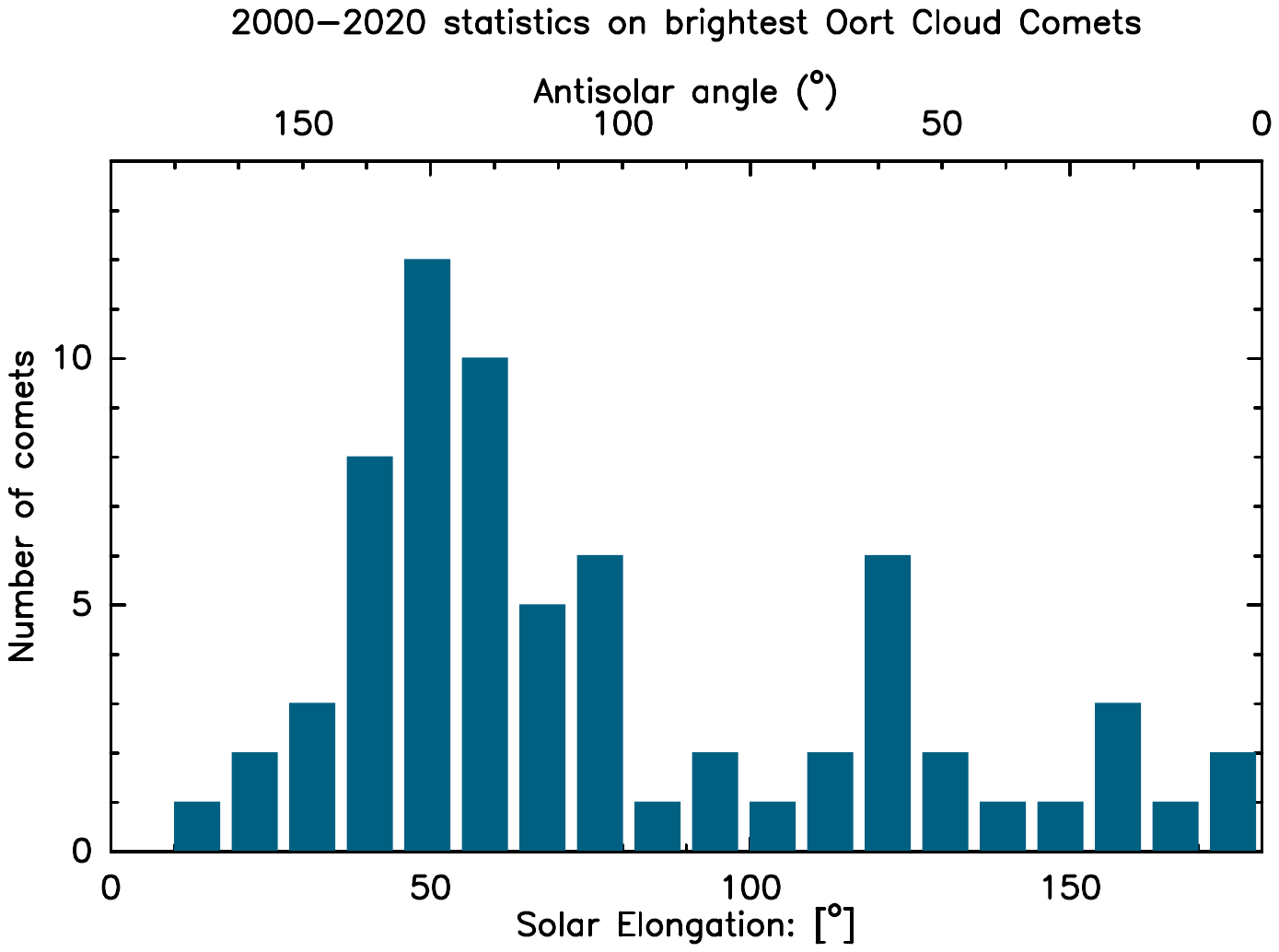}
\end{tabular}
\end{center}
\caption
{ \label{fig:elong} Distribution of solar elongation angles for bright Oort Cloud comets observed in the 2000 -- 2020 period. The elongation shown corresponds to the optimum observing conditions, maximizing the $Q/\Delta$ value. From Ref.\cite{Anderson22}.}
\end{figure}

We note that the telescope Sun avoidance angle was not taken into account in the \emph{Origins} calculations.  Figure~\ref{fig:elong} (from Ref.\cite{Anderson22}) shows the distribution of solar elongation angles of bright Oort Cloud comets observed in the 2000 -- 2020 period.
The elongation shown corresponds to the optimum observing conditions, maximizing the $Q/\Delta$ value, rather than the maximum solar elongation. For the PRIMA Sun avoidance angle of $85^\circ$, only about one third of these comets would be accessible. Applying this factor to all observable comets, we estimate that up to a dozen comets with FOM~$=1 \times 10^{28}$ should be accessible to PRIMA during its 5-year primary mission. The exact number will depend on the actual launch date. In addition, some bright comets may still be accessible at larger elongation angles that do not optimize the sensitivity for HDO detection.

\section{Water isotopic composition as a function of heliocentric distance}\label{sec:rhdep}

Apparitions of bright comets, comparable to comet C/2009 P1 (Garradd), with a water production rate of $2.0 \times 10^{29}$ s$^{-1}$ at 1.8 au, will enable studies of possible variations of the D/H ratio with heliocentric distance. \cmark Such studies have been very limited to date \cite{Biver16, Paganini17, Muller22, Mandt24}, sometimes providing conflicting results.  Future PRIMA \umark observations will determine whether the isotopic composition of water in the coma measured with remote sensing techniques accurately constrains the bulk isotopic composition of the nucleus, or whether isotopic fractionation during sublimation of the nucleus ices plays a role.

Table~\ref{tab:rhdep} gives the expected line flux of the 184.44~$\mu$m HDO line predicted by our non-LTE excitation model in a Garradd-like comet as a function of heliocentric distance. A 20 h observation with FIRESS/FTM corresponds to a 5\,$\sigma$ line flux limit of $4.7 \times 10^{-20}$~W\,m$^{-2}$ (Table~\ref{tab:hdo}). Our computations suggest that the D/H ratio can be measured in such a comet using FIRESS/FTM out to a large heliocentric distance of $\sim 2.7$~au.

\begin{table}
\begin{center}  
\caption{HDO 184.44 $\mu$m line flux as a function of heliocentric distance in a Garradd-like comet.}
\label{tab:rhdep}
\begin{tabular}{cccccccc}
\hline \hline 
  \rule[-3mm]{0mm}{8mm} $r_h$ & $\Delta$ & $T$ & $Q$(H$_2$O) & $\Delta V$ & S(HDO 184 $\mu$m) \\
  \hline 
  1.8      & 1.9 & 47 & $2.0 \times 10^{29}$ & 0.6 & $5.8 \times 10^{-19}$ \\
  2.0      & 2.0 & 40 & $1.6 \times 10^{29}$ & 0.6 & $2.7 \times 10^{-19}$\\
  2.5      & 2.5 & 35 & $1.0 \times 10^{29}$ & 0.5 & $7.8 \times 10^{-20}$ \\
  3.0      & 3.0 & 30 & $0.7 \times 10^{29}$ & 0.5 & $2.0 \times 10^{-20}$ \\
\hline
\end{tabular}
\end{center}
Note: Entries in the table are: heliocentric and geocentric distance (au), coma temperature (K), water production rate (s$^{-1}$), expansion velocity (km\,s$^{-1}$), and the expected line flux of the 184.44~$\mu$m HDO line. 
\end{table}

\section{Summary and conclusions}\label{sec:summ}

Based on our model calculations, we conclude that for a comet with a water production rate of $2 \times 10^{28}$~s$^{-1}$, at a distance $r_h = \Delta =1$, and with a D/H ratio in water 2 times VSMOW (average of past measurements in comets), the 184.44 and 234.64~$\mu$m lines of HDO can be detected at S/N ratio of 5 in $\sim 6$~h observation using FIRESS/FTM. The two lines can be observed simultaneously, together with the 174.37, 181.05, and 183.53 $\mu$m lines of H$_2^{18}$O and the 174.49, 180.33 and 182.09~$\mu$m lines of H$_2^{17}$O in the longest wavelength module of the spectrometer enabling accurate measurements of the D/H ratio.  \cmark The 184.44~$\mu$m HDO, 181.05 H$_2^{18}$O, and 180.33 H$_2^{17}$O lines seem a particularly good combination for isotopic ratio measurements, as our models show no evidence of blending with nearby lines of other molecular species. \umark  Optically thick H$_2^{16}$O lines will also be observed simultaneously, and modeled using a 3D radiative transfer code to better constrain the excitation of water molecules in the coma. Integrations $\sim 4$ times longer should be feasible with FIRESS/FTM, allowing D/H measurements in weaker comets or in comets with a lower D/H ratio, closer to VSMOW.

Our calculations suggest that the 184.44~$\mu$m HDO line is not subject to blending with nearby lines of other abundant coma species. The 234.64~$\mu$m HDO line is close to an excited methanol line, which may have comparable line flux in comets with a very high methanol abundance. However, multiple nearby lines of methanol will allow to model accurately and subtract the possible methanol contribution to the HDO line flux.

The line-to-continuum ratio for detection of these two HDO line at the spectral resolution of FIRESS/FTM is expected to be $\gtrsim 4$~\%. The line-to-continuum ratio is thus not a confounding factor for D/H measurements with PRIMA.

We estimate that up to 36 comets bright enough for D/H measurements will be visible during the nominal 5-year PRIMA primary mission. However, the number of D/H measurements is expected to be reduced by about a factor of 3 due to the constraints imposed by the relatively large Sun avoidance angle of the actively cooled PRIMA telescope. For comparison, only four accurate space-based measurements carried out over the past 25 years. This makes PRIMA a powerful instrument for carrying out comparative isotopic studies between different comet reservoirs, and with inner Solar System measurements in meteorites, as well as for searching for correlations with physical parameters, such as hyperactivity, thus providing quantitative constraints on the dynamical/chemical models of the early Solar System.

Studies of possible variations in the D/H ratio with heliocentric distance in bright comets, comparable to comet C/2009 P1 (Garradd), will be possible with FIRESS/FTM out to $\sim 2.7$~au.  Such observations will determine whether the isotopic composition of water in the coma measured with remote sensing techniques accurately constrains the bulk isotopic composition of the nucleus, or whether isotopic fractionation during sublimation of the nucleus ices plays a role.

\cmark Our sensitivity calculations suggest that a future 2.5-m class FIR space telescope equipped with a state-of-the-art heterodyne receiver targeting isotopic water transitions in the 500~GHz frequency range would be a similarly powerful instrument for studies of isotopic composition of comets. A passively-cooled telescope may offer the advantage of having a smaller Sun avoidance angle compared to PRIMA, enabling observations of comets closer to the Sun, when they are more active and brighter. \umark

\subsection*{Disclosures}
The authors declare that there are no financial interests, commercial affiliations, or other potential conflicts of interest that could have influenced the objectivity of this research or the writing of this paper.

\subsection* {Code, Data, and Materials Availability}
Data tables corresponding to the model spectra shown in Figure~\ref{fig:contam} will be available at the CDS.

\subsection* {Acknowledgments}
This research was carried out at the Jet Propulsion Laboratory, California Institute of Technology, under a contract with the National Aeronautics and Space Administration (80NM0018D0004). MAC was supported by NASA’s Planetary Science Division Internal Scientist Funding Program through the Fundamental Laboratory Research work package (FLaRe). \cmark We thank an anonymous referee for a detailed and helpful report. \umark


\bibliography{biblio}   
\bibliographystyle{spiejour}   


\vspace{2ex}\noindent\textbf{Dariusz C. Lis} is a Scientist at the Jet Propulsion Laboratory, California Institute of Technology.  He joined the submillimeter group at Caltech in 1989, after obtaining his PhD from the University of Massachusetts, Amherst. He was Co-Principal Investigator and Deputy Director of the Caltech Submillimeter Observatory, as well as a member of the US \emph{Herschel} HIFI Science Team. From 2014 to 2019 he was Professor at Sorbonne University in Paris and Director of the Laboratory for Studies of Radiation and Matter in Astrophysics and Atmospheres, a CNRS research laboratory and a scientific department of Paris Observatory. He returned to Caltech, Jet Propulsion Laboratory in 2019. He was a recipient of NASA Group Achievement Awards in 2010 and 2014 and NASA Exceptional Scientific Achievement Medal in 2021. His research interests include solar system small bodies, galactic interstellar medium and star formation, astrochemistry, as well as far-infrared heterodyne instrumentation.
\vspace{1ex}

\vspace{2ex}\noindent\textbf{Martin A. Cordiner} is a Research Associate Professor in Astrochemistry and Planetary Science at Catholic University of America. He obtained a PhD in astrochemistry from The University of Nottingham (UK) in 2006, and now works full-time as a researcher at the NASA Goddard Space Flight Center, in the Astrochemistry Laboratory. He has authored 130 articles on topics from interstellar and circumstellar chemistry to cometary science and planetary atmospheric dynamics, and is principal investigator of the Atacama Large Millimeter/submillimeter Array Large Program “The Large 12P COMA survey: COmetary Molecules with ALMA”.
\vspace{1ex}

\vspace{2ex}\noindent\textbf{Nicolas Biver} holds a permanent research position at the French Center for Scientific Research (CNRS) at Paris Observatory. He received his MS degree in physics from Paris-Cit\'{e} University in 1993, after graduating from the engineering high school “Ecole Centrale” in 1992. He defended his PhD in astrophysics on the observation and modeling of rotational lines of molecules in cometary comae from the Paris-Cit\'{e} University in 1997. He is the author of more than 100 journal papers (25 as first author) and write a book chapter on cometary chemistry (\emph{Comets III}, University of Arizona Press, 2024). His current research interests include the molecular and isotopic composition of cometary volatiles and observations of solar system objects in the millimeter to submillimeter wavelength range.
\vspace{1ex}

\vspace{2ex}\noindent\textbf{Dominique Bockel\'{e}e-Morvan}, Research director at the Observatory of Paris, is working since her PhD on the chemistry of cometary atmospheres through observations in the millimeter and infrared domains, including modeling the  excitation processes in rarefied atmospheres. She was strongly involved in the \emph{Herschel} space telescope, as co-I of the HIFI instrument, and in the \emph{Rosetta} mission, as co-I of the VIRTIS and MIRO instruments. She is now involved in JWST observations of comets and icy satellites.
\vspace{1ex}

\vspace{2ex}\noindent\textbf{Paul F. Goldsmith} received his PhD in 1975 from the University of California Berkeley, after which he was a Member of Technical Staff at Bell Laboratories.  In 1977 he moved to the University of Amherst, Massachusetts, where he became Professor of Physics and Astronomy and Associate Director of the Five College Radio Astronomy Observatory.  In 2003, Goldsmith moved to Cornell University as Professor of Astronomy and Director of the National Astronomy and Ionosphere Center.  In 2005 he moved to the Jet Propulsion Laboratory where he is Senior Research Scientist and Group Supervisor.  Paul Goldsmith’s astronomical research is focused on the structure, formation, and evolution of molecular clouds and their relationship to star formation.  He has carried out a wide variety of ground-based, suborbital and space spectroscopic observations at radio to infrared wavelengths.    Goldsmith was a co-investigator on the Submillimeter Wave Astronomy Satellite (SWAS), launched in 1998. He is the author of the definitive reference text Quasioptical Systems, published in 1998.  At JPL, he has been involved in observations of fine structure line emission and led a large-scale survey of [N{\sc ii}] emission which identified a possible new phase of the interstellar medium which has received the appellation “D-WIM” for Dense, Warm, Ionized Medium.  Goldsmith was NASA Project Scientist for the \emph{Herschel} Space Observatory and used the HIFI instrument on Herschel to make the first definitive multi-transition detection of molecular oxygen in interstellar space.  Currently, he is Project Scientist for GUSTO, a NASA balloon mission that carried out large-scale imaging of the Milky Way in [C{\sc ii}] and [N{\sc ii}] in December 2023 through February 2024. 
\vspace{1ex}

\noindent Biographies and photographs of the other authors are not available.

\listoffigures
\listoftables

\end{spacing}
\end{document}